\documentclass[aps,prl,twocolumn,superscriptaddress,showpacs]{revtex4}
\usepackage[dvips]{graphicx}
\usepackage{subfigure}
\usepackage{multirow}
\begin{document}
\preprint{CU-TP-1193, SHEP 1007}

\title{The $\eta$ and $\eta^\prime$ mesons from Lattice QCD}

\newcommand\riken{RIKEN-BNL Research Center, Brookhaven National
  Laboratory, Upton, NY 11973, USA}
\newcommand\bnl{Brookhaven National Laboratory, Upton, NY 11973, USA}
\newcommand\cu{Physics Department, Columbia University, New York,
  NY 10027, USA}
\newcommand\uconn{Physics Department, University of Connecticut,
  Storrs, CT 06269-3046, USA}
\newcommand\soton{School of Physics and Astronomy, University of
  Southampton,  Southampton SO17 1BJ, UK}
\newcommand\uva{Dept. of Physics, University of Virginia, 
382 McCormick Rd. Charlottesville, VA 22904-4714}

\author{N.H.~Christ}\affiliation{\cu}
\author{C.~Dawson}\affiliation{\uva}
\author{T.~Izubuchi}\affiliation{\bnl}\affiliation{\riken}
\author{C.~Jung}\affiliation{\bnl}
\author{Q.~Liu}\affiliation{\cu}
\author{R.D.~Mawhinney}\affiliation{\cu}
\author{C.T.~Sachrajda}\affiliation{\soton}
\author{A.~Soni}\affiliation{\bnl}
\author{R.~Zhou}\affiliation{\uconn}
\collaboration{RBC and UKQCD Collaborations}

\date{October 8, 2010}

\begin{abstract}
The large mass of the ninth pseudoscalar meson, the $\eta^\prime$, is 
believed to arise from the combined effects of the axial anomaly and the
gauge field topology present in QCD.  We report a realistic, 2+1 flavor, 
lattice QCD calculation of the $\eta$ and $\eta^\prime$ masses and mixing
which confirms this picture.  The physical eigenstates show small
octet-singlet mixing with a mixing angle of $\theta = -14.1(2.8)^\circ$.
Extrapolation to physical light quark mass gives, with statistical errors
only, $m_\eta=573(6)$ MeV and $m_{\eta^\prime}=947(142)$ MeV, consistent 
with the experimental values of 548 MeV and 958 MeV.

\end{abstract}

\maketitle

The relatively large mass of the ninth pseudo-scalar meson, 
the $\eta^\prime$, provides a significant challenge for 
quantum chromodynamics (QCD), the component of the standard
model which describes the interactions of quarks and gluons.
On a naive classical level there are nine conserved axial 
currents.  Given the vacuum breaking of the symmetries which 
these currents generate,  this should imply the existence of 
nine Goldstone bosons, a conclusion inconsistent with the large 
splitting between the 8 octet mesons, $\pi^\pm$, $\pi^0$, 
$K^\pm$, $K^0$, $\overline{K}^0$, $\eta$ and the singlet 
$\eta^\prime$ \cite{Weinberg:1975ui}.  Unique among these nine 
currents, the U(1) axial current, corresponding to the singlet 
$\eta^\prime$ meson, has an anomalous divergence at the quantum 
level.  However, to arbitrary order in perturbation theory this 
anomalous divergence vanishes at zero momentum, continuing to 
imply that the masses of all nine pseudoscalar mesons should 
vanish in the limit of vanishing quark mass.  It is only with 
the discovery of instanton configurations with non-trivial 
topology~\cite{Belavin:1975fg} that a mechanism~\cite{'tHooft:1976up} 
became available that could explain the large $\eta^\prime$ mass.

While these important developments suggest possible consistency 
between QCD and the value of the $\eta^\prime$ mass a direct 
demonstration of the required anomaly-driven, octet-singlet 
splitting has been lacking.  In this paper we present the first 
such demonstration in the realistic case of three light dynamical 
quarks.

The critical role of disconnected diagrams in the study of
the $\eta$ and $\eta^\prime$ and the severe difficulties they 
introduce have been recognized for more than 15 years 
\cite{Kuramashi:1994aj, Venkataraman:1997xi}.   Positivity 
requires the quark propagators that appear in the connected
diagrams to decrease exponentially with increasing time
separation.  For mesons this fall-off roughly matches the
exponential time dependence of the massive, Euclidean-space 
meson propagator and good numerical signals can be seen over a
large range of times.  For terms in which the source and
sink of the meson propagator are not joined by quark propagators,
the needed exponential decrease comes from increasingly large
statistical cancelations implying a rapidly vanishing 
signal-to-noise ratio.  These difficulties have impeded
earlier work~\cite{McNeile:2000hf, DelDebbio:2004ns, Jansen:2008wv, 
Hashimoto:2008xg} on this topic which has employed indirect methods 
or not examined the physical case of up, down and strange 
dynamical quarks; see also Ref.~\cite{Aoki:2006xk}.  

\section{Simulation Details}

Our calculation uses the Iwasaki gauge and domain wall fermion 
actions, a $16^3\times32$ space-time volume with a 
fifth-dimensional extent of 16 and $\beta=2.13$, giving 
an inverse lattice spacing $1/a =1.73(3)$ GeV~\cite{Allton:2008pn}.  
We analyze three ensembles of gauge configurations with light 
sea quark mass $m_l=0.01$, 0.02, 0.03~\cite{Allton:2007hx}.  
(All dimensionful quantities are given in lattice units except 
when physical units are declared.)  These values of $m_l$ yield 
pion masses of 421, 561 and 672 MeV, respectively.  The 0.01 and 
0.02 ensembles were generated using the physical strange quark 
mass $m_s=0.032$~\cite{Mawhinney:2009jy}.  The $m_l= 0.03$ 
ensemble was reported as RHMC II in Ref.~\cite{Allton:2008pn} 
with $m_s=0.04$.  For this ensemble we use reweighting to change 
$m_s$ from 0.04 to 0.032 in 20 mass steps~\cite{Jung:2010jt}. 

We use a Coulomb gauge fixed wall source and sink for the quark 
propagators.  Because of the difficulty of computing the disconnected 
graphs, large statistics are required.  Therefore, we calculate 
propagators for sources on each of our 32 time slices.  The large 
number of Dirac operator inversions ($32\times 12$) that must be 
performed on a single gauge configuration is accelerated by 
computing the Dirac eigenvectors with the smallest 35 ($m_l=0.01$) 
or 25 ($m_l=0.02$, 0.03) eigenvalues and limiting the conjugate 
gradient inversion to the remaining orthogonal subspace.  This 
results in a 60\% speed-up for $m_l=0.01$.  We study 300 
configurations separated by 10 molecular dynamics time units 
for $m_l=0.01$ and 0.02, and 150 configurations separated by 20 
time units for $m_l=0.03$.

We compute four Euclidean space correlation functions between
two pseudoscalar operators $O_l$ and $O_s$:
\begin{equation}
C(t)_{\alpha\beta} = \frac{1}{32} \sum_{t^\prime = 0}^{31} 
    \langle O_\alpha(t+t^\prime)^\dagger O_\beta(t^\prime)\rangle
\;\; \alpha,\beta \in \{l,s\},
\label{eq:corr_fctn}
\end{equation}
summed over the 32 source locations.  Here $O_s=\bar{s}\gamma_5s$ 
and $O_l=(\bar{u}\gamma_5u+\bar{d}\gamma_5d)/\sqrt{2}$, both
SU(2) singlets.

\begin{figure}[t]
  \includegraphics[width=80mm]{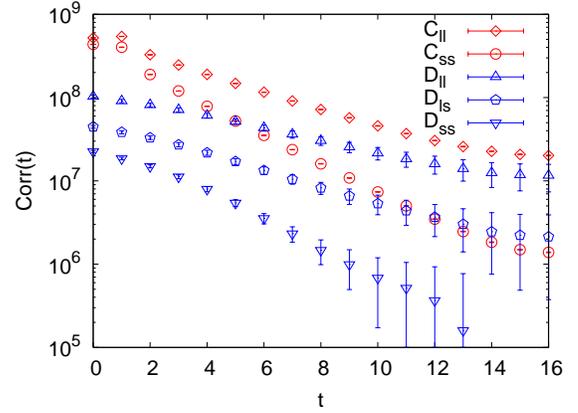}
  \caption{Results for the five contractions which enter the $\eta$-$\eta^\prime$ 
  correlator calculated using the $m_l=0.01$ ensemble.}
  \label{fig:corr}
\end{figure}

The matrix $C(t$) can be expressed in terms of the five amplitudes
represented by the diagrams shown in Fig.~\ref{fig:graph}.  
\begin{equation}
	\left(\begin{array}{cc}
             C_{ll} & C_{ls} \\ 
             C_{sl} & C_{ss}  
	\end{array}\right)
	= \left(\begin{array}{cc}
	{\cal C}_{ll}-2{\cal D}_{ll} & -\sqrt{2}{\cal D}_{ls} \\
	-\sqrt{2}{\cal D}_{sl} & {\cal C}_{ss}-{\cal D}_{ss}
	\end{array}\right).
	\label{eq:OpMatrix}
\end{equation}
This equation shows that neither $O_l$ nor $O_s$ creates an energy 
eigenstate of QCD.  They mix with each other through the disconnected 
diagram ${\cal D}_{sl} = {\cal D}_{ls}$.  The usual expectation that 
such disconnected graphs are small does not apply here.  Figure~\ref{fig:corr} 
shows these amplitudes versus time for the $m_l=0.01$ ensemble.  The 
disconnected graphs decrease more slowly than the connected graphs,
changing the pattern of SU(3) flavor symmetry breaking.

{\renewcommand{\arraystretch}{1.3}
\tabcolsep = 8pt
\arrayrulewidth = 0.02mm
\begin{figure}[t]
  \begin{tabular}{c|c}
  \includegraphics[width=30mm]{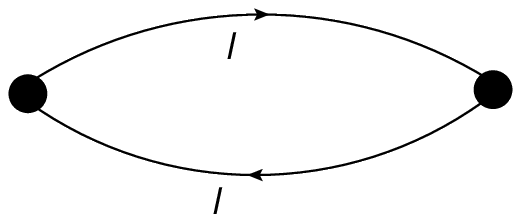}  &
  \includegraphics[width=30mm]{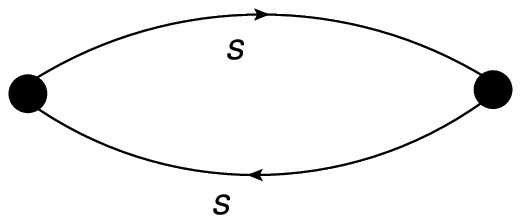} \\
  \hline 
  \rule{0mm}{10mm}
  \includegraphics[width=30mm]{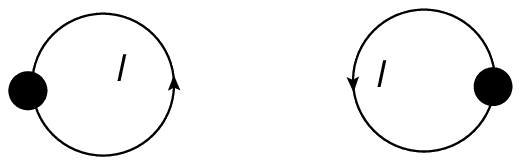}  &
  \includegraphics[width=30mm]{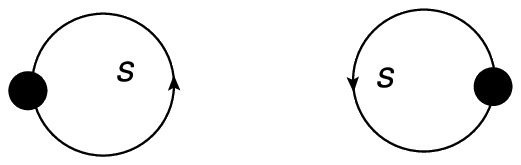}  \\
  \hline 
  \multicolumn{2}{c}{
  \rule{0mm}{10mm}
  \includegraphics[width=30mm]{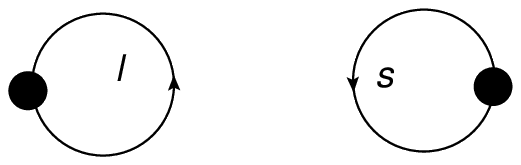} } 
  \end{tabular}
\caption{Five diagrams appearing in the $\eta$ and $\eta^\prime$ correlation 
functions. They are ${\cal C}_{ll}(t)$, ${\cal C}_{ss}(t)$, ${\cal D}_{ll}(t)$, 
${\cal D}_{ss}(t)$, and ${\cal D}_{ls}(t)$ respectively from left to right and 
top to bottom. The solid lines are quark propagators and the solid circles 
$\gamma_5$ insertions.}  \label{fig:graph}
\end{figure}}

Inserting a sum over states into Eq.~\ref{eq:corr_fctn} 
and assuming this sum is dominated by the $\eta$ and $\eta^\prime$ 
for large $t$ we obtain
\begin{equation}
C(t) = A^T D(t) A.
\label{eq:C_matrix}
\end{equation}
where the overlap matrix $A$ is given by:
\begin{equation}
A = \left(\begin{array}{cc}
     \langle\eta|O_l|0\rangle        & \langle\eta|O_s|0\rangle \\
     \langle\eta^\prime|O_l|0\rangle & \langle\eta^\prime|O_s|0\rangle 
     \end{array}\right),
\label{eq:overlap}
\end{equation}
and $D(t)$ is a diagonal matrix with elements $e^{-m_\eta t}$ and 
$e^{-m_{\eta^\prime}t}$.   We chose $A$ real, possible because 
$C(t)$ is real.

Now define a second operator basis with definite SU(3) properties: 
the octet 
$O_8 = (\overline{u}\gamma^5 u+\overline{d}\gamma^5 d-2\overline{s}\gamma^5 s)/\sqrt{6}$ 
and the singlet  
$O_1 = (\overline{u}\gamma^5 u+\overline{d}\gamma^5 d+\overline{s}\gamma^5 s)/\sqrt{3}$.
We will use the Roman indices $a$ and $b$, for these operators, {\it e.g.} 
$\{O_a\}_{a=8,1}$ to distinguish them from the earlier basis 
$\{O_\alpha\}_{\alpha=l,s}$.  Equations analogous to Eqs.~\ref{eq:corr_fctn}, 
\ref{eq:C_matrix} and \ref{eq:overlap} will be obeyed if this second basis with $a,b \in \{ 8,1\}$ is used.

\begin{figure}[t]
  \includegraphics[width=80mm]{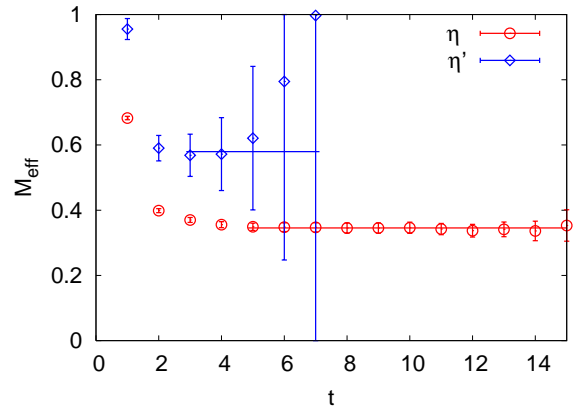}
  \caption{Effective mass plot for the $\eta$ and $\eta^\prime$ states from 
  the $m_l=0.01$ ensemble.}
  \label{fig:effmass}
\end{figure}

We can determine the two masses and the four real elements of the matrix $A$ 
by fitting our data to Eq.~\ref{eq:C_matrix} over an appropriate range of time 
$t$.  To determine this range we examine the product:
\begin{equation}
 C(t_0)^{-1}C(t) = A^{-1}D(t-t_0)A,
\label{eq:eff_mass_gen}
\end{equation}
implying $C(t_0)^{-1}C(t)$ is similar to a diagonal matrix whose 
eigenvalues are exponentials of the masses of interest.  We find the 
best results if $t-t_0$ is large, giving a clean separation of the 
larger, more accurate  $\eta$ eigenvalue and the smaller eigenvalue 
associated with the noisy $\eta^\prime$.  Figure~\ref{fig:effmass} shows 
the eigenvalues obtained from Eq.~\ref{eq:eff_mass_gen}.  Here we plot the 
logarithm of the ratio of each eigenvalue evaluated at $t$ and $t+1$ with
$t_0=2$.  The choice $t_0=2$ and $3 \le t \le 7$ gives 
a recognizable plateau for $m_\eta$ and $m_{\eta^\prime}$.

\section{$\eta - \eta^\prime$ mixing}

It is customary to treat the physical $\eta$ and $\eta^\prime$ states as
mixtures of the pseudo-scalar octet and singlet states which appear in
the SU(3) symmetric limit and to introduce an angle $\theta$ which specifies
this mixing.  In the present calculation we can examine the validity
of this mixing model and attempt to determine $\theta$.  Consider 
the SU(3) symmetric limit $m_l=m_s$ and let $|8\rangle_{\rm sym}$ and 
$|1\rangle_{\rm sym}$ be these lowest energy octet and singlet states
with energies $E_8$ and $E_1$.  We justify this mixing model by assuming 
that when $m_l \ne m_s$ the only important effects are a subset of those implied
by first-order perturbation theory: first-order energy shifts and first-order 
mixing of states but only for those cases enhanced by the relatively small 
energy denominator $E_1 - E_8$.  To zeroth order in $m_s - m_l$ we can write 
$_{\rm sym}\langle a|O_b|0\rangle = Z_a^{1/2} \delta_{ab}$ and we assume 
this relation is unchanged by the first order effects of $m_s - m_l$ 
on the vacuum state --- again neglecting mixing not enhanced by the factor 
$1/(E_1 - E_8)$.  

These assumptions imply that
\begin{equation}
\left(\begin{array}{c} |\eta\rangle \\ |\eta^\prime\rangle \end{array}\right)
 = \left(\begin{array}{cc} \cos(\theta) & -\sin(\theta) \\
                           \sin(\theta) & \cos(\theta)  \end{array}\right)
   \left(\begin{array}{c} |8\rangle_{\rm sym} \\ |1\rangle_{\rm sym} \end{array}\right)
\label{eq:mixing}
\end{equation}
and that the overlap matrix $A$ can be written:
\begin{equation}
A = \left(\begin{array}{cc} 
         Z_8^\frac{1}{2}\cos(\theta) & -Z_1^\frac{1}{2}\sin(\theta) \\
         Z_8^\frac{1}{2}\sin(\theta) &  Z_1^\frac{1}{2}\cos(\theta)
  \end{array}\right),
\end{equation}
for $A$ in the $O_8$ - $O_1$ basis.  The columns of $A$ are thus orthogonal 
and, if $O_8$ and $O_1$ are normalized by multiplication by $Z_8^{-1/2}$ 
and $Z_1^{-1/2}$, the resulting overlap matrix $\hat A$ will be orthogonal.  
Using the results below, we find for the dot product between the columns of 
$\hat A$: -0.016(9) and -0.012(4) for the $m_l=0.01$ and 0.02 ensembles. 

\begin{figure}[t]
  \includegraphics[width=70mm]{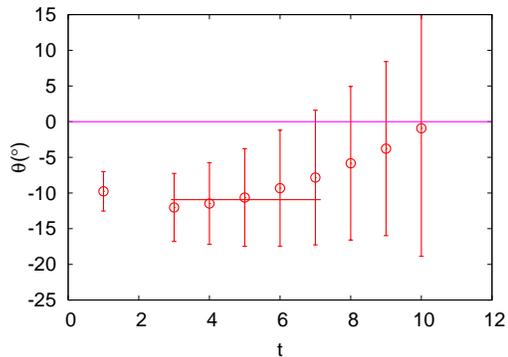}
  \caption{The $\eta-\eta^\prime$ mixing angle $\theta(t)$ determined 
  from Eq.~\ref{eq:eff_mass_gen} for the $m_l=0.01$ ensemble.  While 
  the errors are large, the data is consistent with a single value of 
  about -$10^\circ$ for $3 \le t \le 7$.  (Note, $\theta(t)$ is undefined 
  at $t_0=2$ and off scale at $t=0$.)}  \label{fig:theta_O8_O1}
\end{figure}

We can also extract an effective mixing angle $\theta(t)$ from 
Eq.~\ref{eq:eff_mass_gen}.  This
equation determines each row of $A$ up to an arbitrary constant.  
However, these two undetermined normalization factors as well as 
the factors $Z_8^{1/2}$ and $Z_1^{1/2}$ cancel from the product 
$A_{\eta 1} A_{\eta^\prime 8}/A_{\eta 8} A_{\eta^\prime 1}$, a 
combination which equals $-\tan^2(\theta)$.  The resulting angle 
is shown in Fig.~\ref{fig:theta_O8_O1}.  The small value of $\theta$ 
in the $O_8$ and $O_1$ basis demonstrates the large role played
by the disconnected diagrams.  Had we omitted 
the disconnected diagrams, the matrix $A$ would have been diagonal 
in the $O_l$ and $O_s$ basis giving 
$\sin(\theta) = -\sqrt{2/3}$ or $\theta = -54.7^\circ$, very different
from our $\theta = -14.1(2.8)^\circ$.

\begin{table}
\caption{Meson masses for the $m_l/m_s=0.03/0.04$ ensemble
and at the reweighted value $m_s^{\rm sea}=0.032$ for two
values of the valence strange quark mass $m_s^{\rm val}=0.03$ 
and 0.04.  Here and below only jackknife, statistical errors 
are given.}
\label{tab:reweight}
\begin{ruledtabular}
\begin{tabular}{lllll}
\hline
$m_s^{\rm sea}$        & $m_\pi$ & $m_s^{\rm val}$   & $m_\eta$ & $m_{\eta'}$ \\
\hline
\hline
\multirow{2}{*}{0.04}  & \multirow{2}{*}{0.3907(9)}  & 0.03 & 0.3907(9) & 0.716(49) \\
                       &                             & 0.04 & 0.4316(16) & 0.713(67)\\
\hline
\multirow{2}{*}{0.032} & \multirow{2}{*}{0.3899(11)} & 0.03 & 0.3899(11) & 0.688(60) \\
                       &                             & 0.04 & 0.4328(20) & 0.694(126) \\
\hline
\end{tabular}
\end{ruledtabular}
\end{table}

\begin{table*}
\caption{Masses in lattice units for the nonet of pseudoscalar mesons.}
\label{tab:masses}
\begin{ruledtabular}
\begin{tabular}{llllllll}
\hline
$m_l$(conf)& $m_\pi$    & $m_K$      & $m_\eta$     & $m_{\eta^\prime}$
                                                                & $\theta$ 
                                                                            & $m_\eta$(GMO) \\
\hline
\hline
0.01(300)  & 0.2441(7)  & 0.3272(7)   & 0.3572(24) & 0.600(45) & -$8.3(2.6)^\circ$ & 0.3505(10)\\
0.02(300)  & 0.3251(6)  & 0.3633(6)   & 0.3787(11)  & 0.605(36) & -$5.5(1.4^\circ$) & 0.3752(9)\\
0.03(150)  & 0.3899(11) & \multicolumn{1}{c}{---}
                                      & 0.3988(13)  & 0.689(73) & \multicolumn{1}{c}{---}
                                                                            & \multicolumn{1}{c}{---}\\
\hline
\end{tabular}
\end{ruledtabular}
\end{table*}

\section{Fitting results}

We fit our four correlation functions $C_{ab}(t)$ in two steps.  First, 
using $3 \le t \le 7$ we determine the two masses $m_\eta$
and $m_{\eta^\prime}$ and the four elements of $A$.  Second, 
we fix $A$ to that determined in the first step and fit the 
$\eta\eta$ element of the transformed matrix 
$[(A^T)^{-1}C(t)A^{-1}]_{\eta\eta}$ over the larger range $5 \le t \le 15$ 
to determine more accurately $m_\eta$.  For each fit we minimize 
$\chi^2$ computed from the full covariance matrix, which includes the 
statistical correlations between each measured propagator at each of the 
time separations used.  We treat each configuration as 
independent but check for autocorrelations by grouping the data 
into blocks of size up to 10 and find consistent errors.  As a test for 
long autocorrelations, we compare the first and second halves of our data 
and find consistent results.  Using random sources, 
Refs.~\cite{Gregory:2007ev,Jansen:2008wv} suggest these disconnected 
correlators show large statistical excursions.  We do not see this
behavior.   Our standard wall sources give disconnected and 
connected propagators which follow similar, properly-sampled 
distributions.

For the $m_l=0.03$ ensemble, we reweight the correlation functions 
to change $m_s^{\rm sea}$ from 0.04 to 0.032 and list the 
results in Tab.~\ref{tab:reweight}.   We then linearly interpolate the 
resulting $m_\eta^2$ and $m_{\eta^\prime}$ with strange valence quark 
masses of 0.03 and 0.04 to the point $m_s^{\rm val}=0.032$.  
Table~\ref{tab:masses} lists the resulting masses for the octet states,
$\pi$, $K$, $\eta$, and the singlet state $\eta^\prime$ for each
ensemble.  The final column shows $m_\eta$ determined by the 
Gell-Mann-Okubo(GMO) formula $3m_\eta^2+m_\pi^2=4m_K^2$ using our 
values for $m_\pi$ and $m_K$.  The good agreement with this first order 
formula is consistent with our small octet-singlet mixing.

\begin{figure}[h]
\includegraphics[width=70mm]{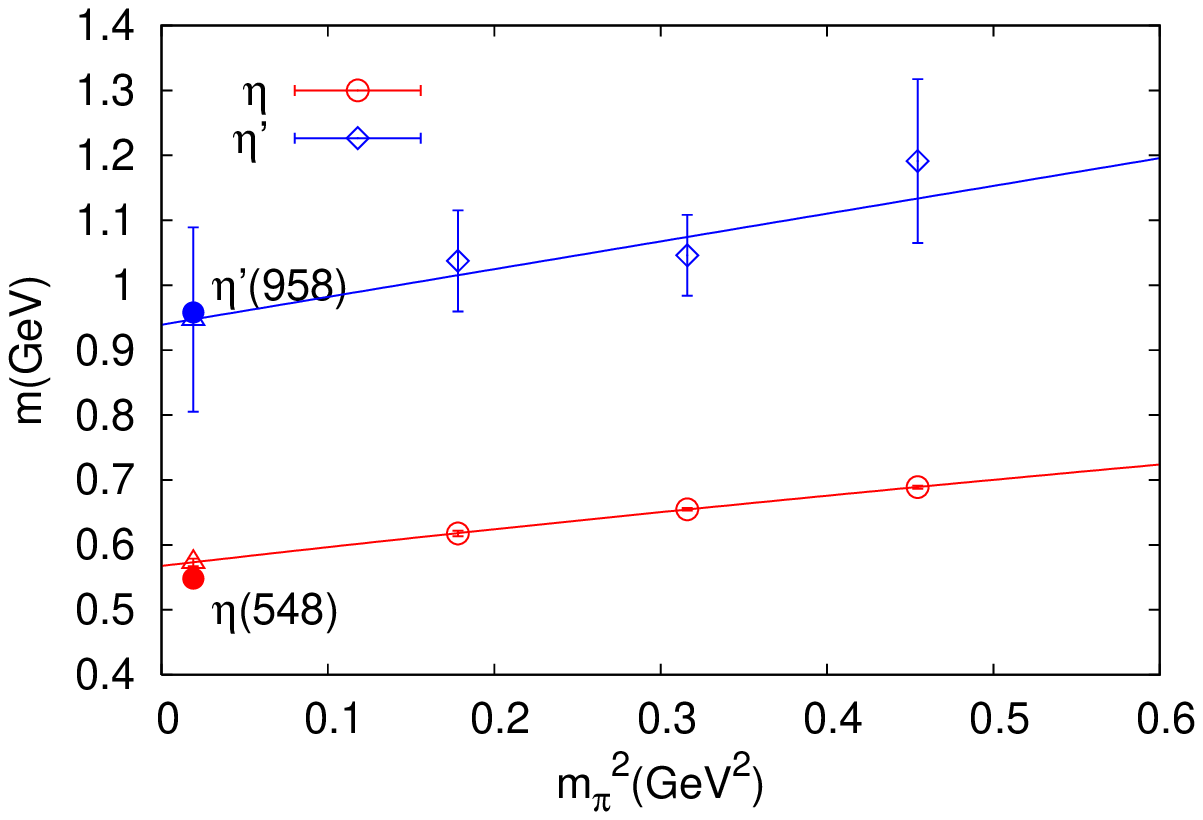} 
\\
\includegraphics[width=75mm]{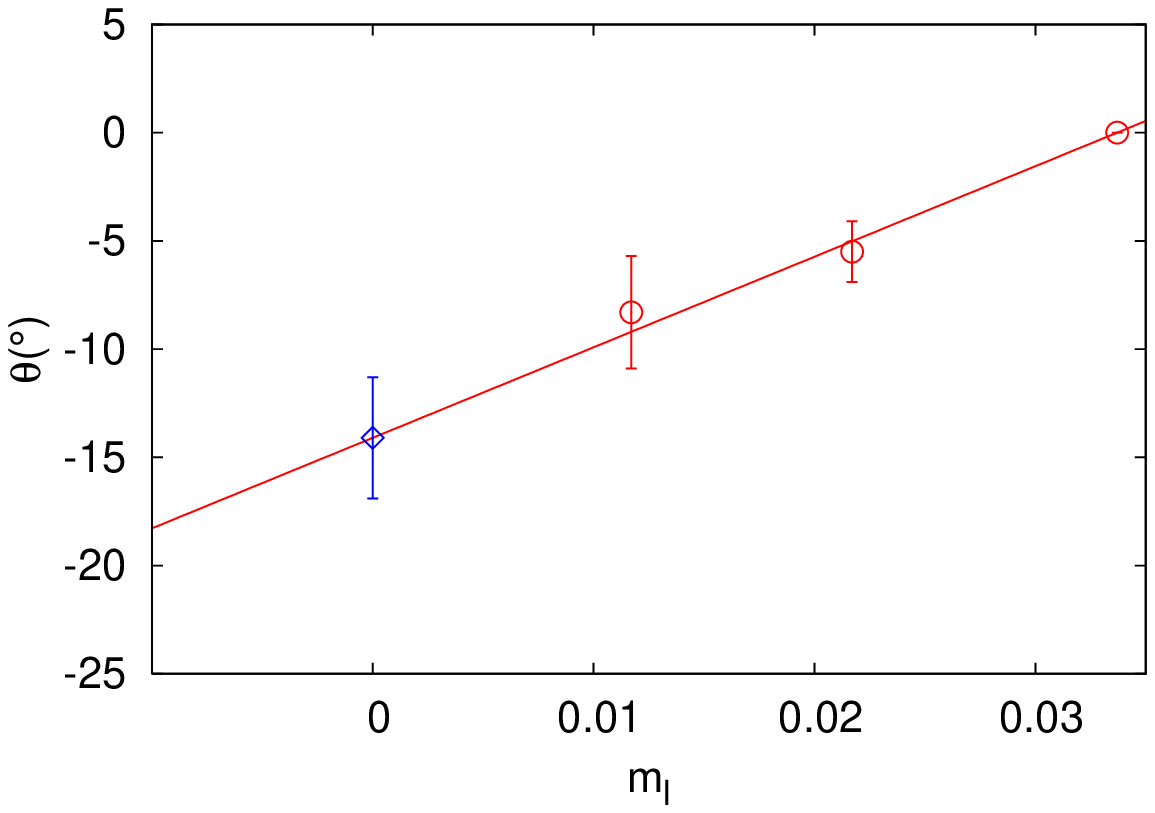} 
\caption{Extrapolation of $m_\eta$, $m_{\eta^\prime}$ (upper)
and $\theta$ (lower) to physical light quark mass (and a negative 
input mass $m_l$).} 
\label{fig:extrapolation}
\end{figure}

In Fig.~\ref{fig:extrapolation} we show a linear extrapolation of
$m_{\eta^\prime}$ and $m_\eta^2$ as a function of $m_\pi^2$ to the 
physical value of $m_\pi$, consistent with NLO chiral perturbation 
theory.  (Note, the curvature of the $m_\eta$ fit is barely visible.)  
We find $m_\eta=573(6)$ MeV and $m_{\eta^\prime}=947(142)$ MeV, where 
the errors are statistical.   To verify our choice of $m_s$, 
we extrapolate the kaon mass and find the physically consistent 
value 497.4(7) MeV.  Also shown is a similar linear extrapolation for 
$\theta$ giving $\theta = -14.1(2.8)^\circ$, in agreement with 
the range $-10^\circ$ to $-20^\circ$ of phenomenological 
values~\cite{Feldmann:1999uf}.

We have described a 2+1 flavor calculation of the masses
and mixing for the $\eta$ and $\eta^\prime$ mesons finding results 
agreeing within their 15\% error with experiment.  The near orthogonality of
the mixing matrix $\hat A$ is consistent with physical states 
which are simple mixtures of SU(3) octet and singlet states.  
Given our large statistical errors we have not analyzed the smaller 
systematic errors arising from our single lattice spacing, large 
light quark masses and finite volume which other 
calculations~\cite{Allton:2008pn,Kelly:2009fp,Mawhinney:2009jy} suggest 
are $\approx 4\%$, 5\% and 1\%.  However, to this 
accuracy our calculation demonstrates that QCD can explain the 
large mass of the ninth pseudoscalar meson and its small 
mixing with the SU(3) octet state.

We thank our RBC/UKQCD collaborators for many helpful ideas and 
BNL, the University of Edinburgh, PPARC and RIKEN for providing 
the facilities on which this work was performed.  This work 
was supported by STFC Grant ST/G000557/1, EU contract 
MRTN-CT-2006-035482(Flavianet), U.S.~DOE grant/contract numbers 
DE-AC02-98CH10886, DE-FG02-92ER40699 and JSPS Grant-in-Aid numbers 
19740134, 22540301.

\bibliography{citations}

\end{document}